\documentclass[lettersize,journal]{IEEEtran}
\usepackage{amsmath,amsfonts}
\usepackage{algorithmic}
\usepackage{algorithm}
\usepackage{array}
\usepackage[caption=false,font=normalsize,labelfont=sf,textfont=sf]{subfig}
\usepackage{textcomp}
\usepackage{stfloats}
\usepackage{url}
\usepackage{verbatim}
\usepackage{graphicx}
\usepackage{cite}
\usepackage{multirow}
\usepackage{booktabs}
\usepackage{siunitx}
\usepackage{booktabs}   
\usepackage{arydshln}
\usepackage{tablefootnote}
\usepackage[hidelinks]{hyperref}
\newenvironment{DIFnomarkup}{}{}

\begin{document}


\title{Noise-Aware Lombard Speech Synthesis via Flow-GRPO for Intelligibility Optimization}

\author{Zhaoyang Wang, Dingding Yao, Chengzhong Wang, and Junfeng Li
\thanks{This research is partially supported by National Natural Science Foundation of China (No. 12425413, 12574509), Beijing Natural Science Foundation (L257001, L247022).}
\thanks{Zhaoyang Wang, Dingding Yao, Chengzhong Wang and Junfeng Li are with the Key Laboratory of Speech Acoustics and Content Understanding, Institute of Acoustics, Chinese Academy of Sciences, Beijing 100190, China, and also with University of Chinese Academy of Sciences, Beijing 100049, China. (Email: wangchaoyang@hccl.ioa.ac.cn, yaodingding@hccl.ioa.ac.cn, wangchengzhong@hccl.ioa.ac.cn, lijunfeng@hccl.ioa.ac.cn)}
\thanks{Corresponding Author: Junfeng Li.}}

\markboth{Journal of \LaTeX\ Class Files,~Vol.~14, No.~8, August~2021}%
{Shell \MakeLowercase{\textit{et al.}}: Lombard Speech Synthesis Based on GRPO for Higher Intelligence in Noise Environments}


\maketitle

\begin{abstract}
Voice-based conversational agents are increasingly deployed in real-world settings, where environmental noise often degrades the interactive experience. Therefore, robust acoustic adaptability is essential to maintain natural and intelligible human-computer communication.
However, most existing systems remain acoustically insensitive and fail to emulate the human Lombard effect, in which speakers involuntarily adjust their vocal production to preserve intelligibility in the presence of noise. As a result, their communicative reliability degrades under adverse acoustic conditions.
To remain intelligible in noisy environments, conversational agents require a mechanism to emulate noise-induced adaptive behavior during synthesis.
We propose NIL-TTS, a noise-induced Lombard TTS framework. We integrate a feature-wise linear modulation (FiLM) mechanism into a flow-matching backbone for dynamic vocal adaptation. A two-stage training paradigm is utilized: initial supervised fine-tuning on paired Lombard corpora, followed by Flow-GRPO post-training to optimize for intelligibility. 
Evaluations confirm that NIL-TTS autonomously adjusts its output in response to varying noise conditions, exhibiting key correlates of the human Lombard effect. 
Driven by these adaptive modulations, the synthesized speech achieves significant intelligibility improvements that scale proportionally with noise intensity. 
Ultimately, this work presents a flow-matching and reinforcement learning-based advancement within the adaptive Lombard speech synthesis paradigm for challenging real-world scenarios.
\end{abstract}


\section{Introduction} 
\IEEEPARstart{T}{he} emergence of high-fidelity speech synthesis with natural prosody has accelerated the development of voice-based conversational agents (VCAs).
Driven by advances in speech-language models (SLMs), these agents are transitioning toward increasingly lifelike interactive dialogues \cite{guo2025recent, cui2025recent}. Consequently, VCAs are evolving beyond static digital assistants into embodied companions deployed in complex physical environments, ranging from in-vehicle settings to noisy public spaces \cite{weng2016conversational}. This paradigm shift has rendered the need for environmental adaptability in VCAs increasingly critical.

Despite strong linguistic capabilities exhibited by modern large language models (LLMs), most existing conversational agents remain acoustically insensitive to their surroundings. In contrast, human communicative behavior is shaped not only by conversational partners but also by the surrounding acoustic context \cite{garnier2010influence}. A typical example is the Lombard effect—a physiological feedback loop wherein speakers involuntarily adjust their voice to preserve intelligibility in the presence of noise \cite{lombard1911signe, brumm2011evolution, luo2018lombard}. This adaptive mechanism involves coordinated acoustic and phonetic changes, including a higher fundamental frequency ($F_0$), longer vowel durations, shifted formant frequencies, and a flatter spectral tilt \cite{lu2008speech, cooke2010spectral}. This process reflects an adaptive communicative strategy that ensures robust information transmission \cite{tartter1993some, castro2024modeling}. 

Due to the lack of such environment-driven regulatory mechanisms, current VCAs suffer substantial degradation in intelligibility during real-world deployment.
To address this limitation, recent studies have explored how to integrate Lombard characteristics into Text-to-Speech (TTS) systems, investigating methodologies such as SNR-feedback-based adaptation, online Lombard adaptation, and controllable vocal-effort synthesis \cite{lopez2014speech, rottschafer2015online, novitasari2021dynamically, novitasari2022machine, bollepalli2019lombard, paul2020enhancing, hu2021whispered, raitio2022vocal, akti2026lombard}.
Existing paradigms predominantly hinge on three strategies: supervised fine-tuning utilizing dedicated Lombard corpora \cite{bollepalli2019lombard, hu2021whispered}, heuristic loudness adjustment modulated by Signal-to-Noise Ratio (SNR) feedback \cite{novitasari2021dynamically}, and speaker-adaptive Lombard style embedding \cite{lobato2025gradual, akti2026lombard}. However, these approaches exhibit inherent constraints. First, the scarcity of high-quality Lombard data often leads to rigid and static acoustic mappings. As a result, models may manipulate the spectral structure in a mechanistic fashion, neglecting natural biological variability and exhibiting poor generalization to unseen speakers. Second, conventional supervised learning and speaker-transition methodologies intrinsically optimize for acoustic mimicry (i.e., $L_2$ similarity to training data), while intelligibility—the goal of the Lombard effect—is rarely posited as a direct optimization objective. Finally, the Lombard effect is not a static style but a behavior shaped by environmental interactions, acutely sensitive to the spectral properties and intensity of ambient noise \cite{junqua1996influence, summers1988effects}. 
Thus, the above approaches still cannot model environment-dependent dynamics under real-world noise conditions.

Recently, Reinforcement Learning (RL) has emerged as an effective post-training paradigm for aligning generative models with complex human preferences and downstream objectives \cite{gao2025differentiable, sun2025f5r}. Yet, its potential for Lombard speech synthesis remains unexploited. RL optimizes its policy through continuous environmental interaction \cite{zhang2025survey}. This process resembles the human auditory-vocal feedback loop, where speakers dynamically adjust their voices to counter ambient noise. 

Motivated by this mechanistic similarity between biological adaptation and RL, we introduce Noise-Induced Lombard TTS (NIL-TTS), the first RL-driven framework designed explicitly for noise-aware, high-intelligibility speech synthesis. Specifically, we augment a foundational flow-matching TTS model with a FiLM-based environmental sensing pathway. 
To move beyond acoustic mimicry, we propose a two-stage training paradigm: initial supervised fine-tuning (SFT) to establish foundational style mappings, followed by an online Flow-GRPO post-training stage that directly optimizes for intelligibility via ASR-derived rewards. Audio samples are available on our demo page\footnote{\url{https://kalenon.github.io/NIL-TTS-demo/}}. Our key contributions are summarized as follows:

1) We propose a novel environment-driven synthesis framework (NIL-TTS) that equips a TTS system with the capability for autonomous acoustic adaptation to ambient noise.

2) We empirically validate the mechanistic synergy between reinforcement learning and the biological Lombard effect, demonstrating RL's efficacy in achieving fine-grained control over prosodic and acoustic compensatory strategies.

3) We demonstrate that NIL-TTS discovers and implements complex acoustic adjustments under adverse conditions. These environment-induced shifts yield substantial improvements in intelligibility, marking a critical advancement toward truly situated, embodied conversational intelligence.

\section{Related Work}
\subsection{F5-TTS}
F5-TTS is a non-autoregressive (NAR) text-to-speech (TTS) system based on flow matching with a Diffusion Transformer (DiT) backbone \cite{chen2025f5}. It introduces a simplified synthesis paradigm that bypasses the complex auxiliary modules common in earlier NAR systems, such as duration models, separate text encoders, and explicit phoneme-level alignment. Instead of rigid alignment, F5-TTS refines text representations using ConvNeXt V2 blocks \cite{woo2023convnext} and pads character sequences with filler tokens to match target speech lengths, allowing the model to learn prosody and rhythmic contours implicitly during the iterative denoising process.

During inference, F5-TTS employs Sway Sampling to allocate the Number of Function Evaluations (NFE) non-uniformly across the flow trajectory. This strategy places greater emphasis on early generation stages, which are critical for establishing the global acoustic structure of the utterance. 
Empirically, the model supports high-fidelity zero-shot synthesis. 
The architectural simplicity provides a flexible foundation for the environment-aware Lombard adaptation tasks investigated in this work.

\subsection{FiLM}
Feature-wise Linear Modulation (FiLM) is a general conditioning method that influences neural network computation through feature-wise affine transformations \cite{perez2018film}. By scaling and shifting activations based on external signals, FiLM can selectively emphasize or suppress specific feature channels, providing a highly efficient mechanism for injecting external control into deep architectures \cite{birnbaum2019temporal}. 
In Lombard speech synthesis, FiLM provides a suitable pathway for conditioning speech production on the acoustic environment \cite{akti2026lombard}. 
Its architecture enables it to serve as a scalable bridge between ambient noise perception and the resulting vocal adaptations, allowing the model to adjust its generative trajectory within the Diffusion Transformer backbone.

\subsection{Group Relative Policy Optimization (GRPO)}
Group Relative Policy Optimization (GRPO) is an efficient reinforcement learning algorithm first used to improve reasoning in large language models (LLMs) \cite{shao2024deepseekmath}. Compared with Proximal Policy Optimization (PPO) \cite{schulman2017proximal}, GRPO avoids training a separate critic model of similar scale to the policy model, which reduces memory and computation costs \cite{zhang2025survey}.

In GRPO, the policy generates a group of outputs for each prompt. The advantage of each output is then computed from its reward relative to the group mean and standard deviation. This relative formulation matches the comparative nature of many reward signals and improves training efficiency. 
In standard GRPO formulations, a Kullback-Leibler (KL) penalty can be introduced to constrain policy updates relative to a reference model.

Recently, a related framework integrates GRPO into a flow-matching TTS architecture by reformulating deterministic flow outputs into probabilistic Gaussian distributions \cite{sun2025f5r}, demonstrating the efficacy of GRPO for general speech synthesis. 

\subsection{Flow-GRPO}
Flow-GRPO \cite{liu2026flow} extends GRPO to flow-matching models in continuous time. In speech generation, many RL methods focus on autoregressive models or discrete token spaces. In contrast, Flow-GRPO enables online RL directly in continuous mel-spectrogram space. It samples a group of speech trajectories for each prompt and computes relative advantages from group-normalized rewards.

For flow-matching architectures, this design allows direct optimization of the vector field $v_t$ with respect to human-centered quality metrics, without introducing a critic model. Through iterative updates, Flow-GRPO can refine generation trajectories and improve speech quality beyond standard supervised fine-tuning. 


\section{Methodology}
This section details the architecture and training paradigm of NIL-TTS. We build upon a continuous flow-matching TTS architecture, extending it with two core mechanisms. First, we introduce an environmental acoustic sensing pathway via Feature-wise Linear Modulation (FiLM) that encodes ambient acoustic conditions. Second, to transcend the limitations of purely supervised acoustic mimicry, we propose a two-stage training paradigm: an initial supervised fine-tuning (SFT) stage, followed by a Flow-GRPO post-training stage. This hybrid approach ensures that the model not only learns the foundational acoustic priors of Lombard speech but also explicitly optimizes its vocalization strategies to maximize intelligibility through continuous environmental interaction.

\subsection{Architectural Framework}

\subsubsection{Base Flow-Matching Acoustic Model}
The foundation of our system is a conditional flow-matching (CFM) acoustic model that generates mel-spectrograms from text. Let $x_1$ represent the target mel-spectrogram and $x_0 \sim \mathcal{N}(0, I)$ be a sample from the prior noise distribution. The model learns a parameterized vector field $v_\theta(x_t, t, c)$ that maps the noise $x_0$ to the target speech $x_1$ along a probability path $p_t(x)$, where $t \in [0, 1]$ is the time step. The conditioning information $c = (c_{text}, c_{ref})$ includes linguistic features from the input text and speaker identity features from a reference audio prompt. The model is trained by minimizing the flow-matching objective:

\begin{equation}
\label{CFM}
\mathcal{L}_{CFM} = \mathbb{E}_{t, x_0, x_1} \left[ \left\| v_\theta(\phi_t(x_0, x_1), t, c) - (x_1 - x_0) \right\|^2 \right]
\end{equation}
where $\phi_t(x_0, x_1) = (1-t)x_0 + t x_1$ defines a linear interpolation between the prior noise and the data. The vector field $v_\theta$ is parameterized by a Diffusion Transformer (DiT), which serves as the core of our generative backbone.

\subsubsection{Environment-Aware Injection via FiLM}
\begin{figure*}[thb]
\centering
\includegraphics[width=\textwidth]{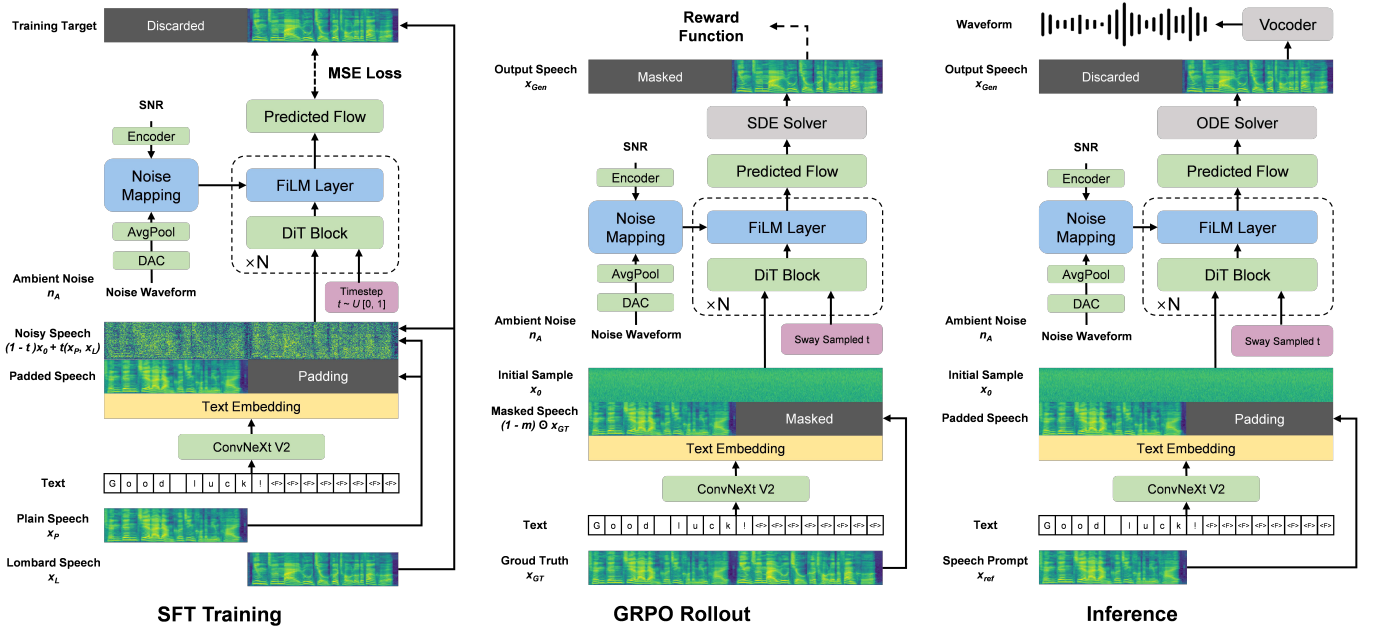}
\caption{An overview of the NIL-TTS pipeline. Left: The supervised fine-tuning (SFT) stage, where the model is trained with a conditional flow matching loss on paired plain and Lombard speech. Middle: The Flow-GRPO rollout process, which leverages an SDE solver to generate a group of exploratory sample trajectories for policy optimization. Right: The inference process, which uses an ODE solver for deterministic speech generation.}
\label{fig:pipeline}
\end{figure*}

Standard DiT architectures typically lack explicit mechanisms for environmental feedback.
To synthesize situated Lombard speech, the model should first perceive the ambient acoustic stimuli and modulate its generative trajectory accordingly. We achieve this by explicitly conditioning the DiT on environmental information using a Feature-wise Linear Modulation (FiLM) mechanism.

The ambient environment is characterized by two distinct components: the normalized noise waveform and its corresponding Signal-to-Noise Ratio (SNR) with speech. 
We employ a pre-trained noise encoder to extract rich acoustic embeddings $E_{noise} \in \mathbb{R}^{T \times D_{enc}}$ from the noise waveform. 
Concurrently, the scalar SNR is projected into a vector representation $E_{snr}$ using a multilayer perceptron (MLP). These representations are fused to construct a global environmental condition vector $c_{env}$:

\begin{equation}
\label{E_noise}
c_{env} = \text{MLP}_{proj}\left( [ \text{AvgPool}(E_{noise}) \parallel E_{snr} ] \right)
\end{equation}

The average pooling over the noise embeddings ($\text{AvgPool}$) removes the frame-level temporal dynamics of the noise, producing a global, utterance-level environmental condition. This global context $c_{env}$ is injected directly into the hidden states of the DiT blocks. Specifically, for the $i$-th Transformer block, 
the hidden state \(h_i\) (following the AdaLN conditioning and self-attention mechanism) is modulated by the FiLM parameters: 

\begin{equation}
\label{film_con}
\gamma_i, \beta_i = W_{i}(c_{env})
\end{equation}

\begin{equation}
\label{film_apply}
\tilde{h}_i = h_i \odot \gamma_i + \beta_i
\end{equation}
where $W_{i}$ is a linear projection layer generating distinct scaling ($\gamma_i$) and shifting ($\beta_i$) parameters for each specific block $i$, and $\odot$ denotes element-wise multiplication. This modulation pathway establishes the vital sensory link, allowing external acoustic conditions to continuously govern the internal speech generation process.
Furthermore, to preserve the foundational phonetic alignment learned from clean speech, this FiLM-based modulation is applied exclusively to the deeper transformer blocks ($i > 2$).

\subsection{Two-Stage Training Paradigm}
\subsubsection{Supervised Fine-Tuning (SFT)}
In the initial training stage, the environment-aware DiT model is fine-tuned on a corpus of paired plain and Lombard speech samples. As illustrated in Fig.~\ref{fig:pipeline} (left), the mel-spectrograms of plain speech ($x_P$) and Lombard speech ($x_L$) from the same speaker are concatenated along the time dimension, forming a composite target $(x_P, x_L)$. The model is conditioned on the input text, a padded version of the plain speech $(x_P, \text{padding})$, and the ambient noise condition $n_A$.

The model is optimized to minimize the masked flow-matching loss $\mathcal{L}_{CFM}$ (Eq.~\ref{CFM}). This supervised stage equips the model with a foundational heuristic for mapping ambient noise conditions to basic Lombard acoustic features, serving as the baseline policy for subsequent interactive refinement.


\begin{figure*}[!t]
\centering
\includegraphics[width=0.7\textwidth]{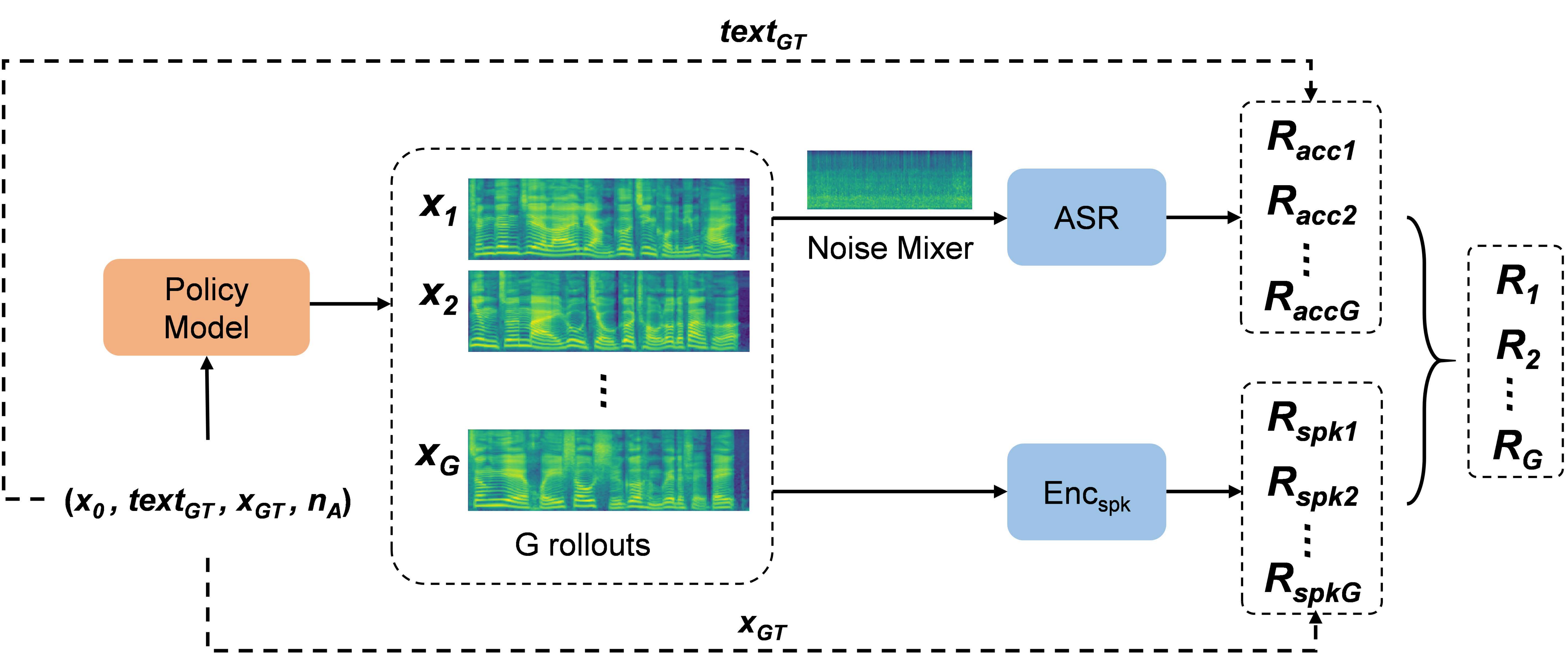}
\caption{Reward computation in Flow-GRPO. The objective function acts as a proxy for human auditory feedback. The intelligibility reward ($R_{acc}$) mixes rollouts with ambient noise before ASR transcription to evaluate robustness. The speaker similarity reward ($R_{spk}$) utilizes a speaker encoder to penalize identity distortion against the ground truth ($x_{GT}$).}
\label{fig:reward}
\end{figure*}

\subsubsection{Flow-GRPO Post-Training}
While SFT imparts the basic acoustic patterns of Lombard speech, its objective primarily optimizes data likelihood rather than intelligibility. However, the model is not explicitly incentivized to mitigate environment-induced intelligibility degradation. To bridge this gap, we introduce a reinforcement learning stage that adapts Flow-GRPO to flow-matching trajectories \cite{liu2026flow}.

A fundamental challenge in applying RL to continuous flow models is that standard generation relies on a deterministic Ordinary Differential Equation (ODE), which lacks the stochasticity requisite for RL exploration:

\begin{equation}
\label{ODE}
\mathrm{d} x_t = v_t(x_t) \mathrm{d} t
\end{equation}

Acoustic exploration serves as a computational analogue of human vocal trial-and-error processes. To enable this, following the methodology established in Flow-GRPO \cite{liu2026flow}, we reformulate the generation process into a Stochastic Differential Equation (SDE). This stochastic framework facilitates exploration while rigorously preserving the marginal distributions of the underlying ODE.

\begin{equation}
    \label{flow sde}
    \mathrm{d} x_t = \left[ v_t(x_t) + \frac{\sigma_t^2}{2t} (x_t + (1 - t) v_{t}(x_t)) \right] \mathrm{d} t + \sigma_t \mathrm{d} \omega
\end{equation}
where $\mathrm{d} \omega$ is the standard Wiener process, and $\sigma_t$ is a diffusion coefficient regulating stochasticity. Discretized via the Euler-Maruyama method, the update rule becomes:

\begin{equation}
\begin{split}
\label{flow sde update}
 x_{t+\Delta t} = x_t &+ \left[ v_\theta(x_t, t) + \frac{\sigma_t^2}{2t} (x_t + (1 - t) v_\theta(x_t, t)) \right] \Delta t \\
 &+ \sigma_t \sqrt{\Delta t} \epsilon 
\end{split}
\end{equation}
where $\epsilon \sim \mathcal{N}(0, \boldsymbol{I})$ introduces the noise necessary for generating diverse vocal trajectories. Following \cite{liu2026flow}, we set $\sigma_t = a\sqrt{t/(1-t)}$, with hyperparameter $a$ dictating the exploration boundary.

In this post-training rollout phase (Fig.~\ref{fig:pipeline}, middle), the ground-truth mel-spectrogram $x_{GT}$—which denotes the plain reference speech, rather than a noise-corrupted or Lombard ground truth—is randomly masked to yield a partially observed target $(1-m)\odot x_{GT}$. A noise sample $x_0 \sim \mathcal{N}(0, I)$ is drawn, and the model is optimized as an active policy $\pi_{\theta}$. For each input condition, the policy generates a group of $G$ sample trajectories $\{ \tau_1, \dots, \tau_G \}$ using the SDE solver.

The reward function $\mathcal{R}$ (Fig.~\ref{fig:reward}) is engineered to computationally simulate human auditory judgment under noise \cite{sun2025f5r}:

\begin{equation}
\label{reward}
\mathcal{R} = \lambda_{spk} R_{spk} + \lambda_{acc} R_{acc}
\end{equation}
where $\lambda_{spk}$ and $\lambda_{acc}$ balance identity preservation against intelligibility constraints.

\begin{itemize}
\item \textbf{Intelligibility Reward ($R_{acc}$):} The generated speech is mixed with the target noise at a specified SNR and transcribed by an external Automatic Speech Recognition (ASR) model. We define this reward as:

\begin{equation}
R_{acc} = 1 - \mathrm{WER}(text_{rollout}, text_{GT})
\end{equation}


In this setup, the ASR acts as a proxy for human auditory feedback. This feedback encourages the policy to adapt to the ambient noise. Consequently, the model discovers acoustic strategies that maximize transcription accuracy.

\item \textbf{Speaker Similarity Reward ($R_{spk}$):} The cosine similarity between the generated speech's speaker embedding and that of the ground-truth reference is measured:

\begin{equation}
R_{spk} = \mathrm{SIM}(emb_{rollout}, emb_{GT})
\end{equation}

This intrinsic penalty ensures that the intense vocal adaptations required for intelligibility do not distort the original speaker identity.
\end{itemize}

Advantage estimation is executed via group-level normalization:

\begin{equation}
\label{advantage}
A_i = \frac{\mathcal{R}_i - \mu_{\mathcal{R}}}{\sigma_{\mathcal{R}}}
\end{equation}
where $\mu_{\mathcal{R}}$ and $\sigma_{\mathcal{R}}$ represent the group's reward mean and standard deviation. The policy is updated by maximizing a clipped surrogate objective, penalized by the Kullback-Leibler (KL) divergence to anchor the policy against catastrophic drift from the SFT reference model $\pi_{ref}$:

\begin{equation}
\label{grpo-loss}
\begin{split}
\mathcal{L}_{GRPO} = &- \frac{1}{G} \sum_{i=1}^G \min\left( r_i(\theta) A_i, \text{clip}(r_i(\theta), 1-\epsilon, 1+\epsilon) A_i \right) \\
& - \beta \mathbb{D}_{KL}(\pi_\theta \| \pi_{ref})  
\end{split}
\end{equation}
where $r_i(\theta) = \frac{\pi_\theta(\tau_i)}{\pi_{old}(\tau_i)}$ denotes the importance sampling ratio.

\subsection{Inference}
During inference, the goal is to generate a deterministic and highly intelligible sample of Lombard speech for a specified text and noise condition. Conditioned on a target speaker reference $x_{ref}$, generation proceeds via the deterministic ODE solver (Eq.~\ref{ODE}) using a fixed number of integration steps (e.g., 32 NFE). This process bypasses SDE stochasticity to ensure stable and highly reproducible output (Fig.~\ref{fig:pipeline}, right). Finally, a pre-trained vocoder synthesizes the predicted mel-spectrogram into the final acoustic waveform.

\section{Experiments}
\subsection{Datasets}
The proposed two-stage training framework leverages distinct datasets to ensure both stylistic fidelity and cross-domain generalization.

\subsubsection{Supervised Fine-Tuning (SFT) Dataset}
The initial stage utilizes the Chinese Enhanced Mandarin Lombard Grid (EMALG) corpus \cite{li2024emalg} and the English LombardGrid corpus \cite{alghamdi2018corpus}. EMALG consists of 10,200 meaningful utterances recorded by 34 native Mandarin speakers. It includes a plain reference style recorded at 40 dB SPL and two Lombard styles induced at 55 dB (L1) and 80 dB (L2). The LombardGrid dataset comprises 5,400 utterances from 54 speakers, featuring paired plain speech at 30 dB SPL and Lombard speech at 80 dB SPL. These datasets provide the foundational high-fidelity paired data required to establish the baseline mapping between noise and acoustic adaptation.

To simulate diverse real-world acoustics, we augment the training signals with environmental noise from the DEMAND database \cite{thiemann2013diverse}. This database provides recordings across six broad categories: Domestic, Office, Public, Transportation, Nature, and Street. We utilize DEMAND in this initial stage because its categorical environments are well-suited for establishing foundational noise-to-acoustic mappings.

\subsubsection{Flow-GRPO Post-Training Dataset}
To facilitate stochastic exploration of the acoustic state space and promote generalization to naturalistic conversational speech, the reinforcement learning phase exclusively utilizes the Mandarin WenetSpeech4TTS-Premium subset \cite{ma2024wenetspeech4tts}. This corpus comprises 945 hours of high-fidelity Mandarin speech, specifically filtered to ensure superior perceptual quality with DNSMOS \cite{reddy2021dnsmos} scores exceeding $4.0$. The multi-domain nature of this data (e.g., audiobooks, interviews, and podcasts) provides the rich linguistic and prosodic variability necessary for effective policy optimization.

For environmental simulation during post-training, we employ the DNS Challenge (ICASSP 2021) noise corpus \cite{reddy2021icassp}, which encompasses approximately 181 hours of diverse noise samples across 150 distinct audio classes, including office, transportation, and social environments. 
Exposing the model to this massive diversity of acoustic interference forces the policy to develop adaptive strategies rather than overfitting to specific noise profiles.

\subsubsection{Evaluation Dataset}
To evaluate the effectiveness of the proposed model in zero-shot speech synthesis, we conduct objective evaluations using the public benchmark configuration of the Seed-TTS objective test set \cite{anastassiou2024seed}. This includes 2,020 Mandarin utterances derived from the DiDiSpeech corpus \cite{guo2021didispeech} and 1,088 English utterances extracted from the Common Voice dataset \cite{ardila2020common}.

To assess environmental adaptability under adverse conditions, these clean samples are systematically perturbed with a representative suite of noise profiles from the NoiseX-92 database \cite{varga1993assessment}. This corpus is specifically curated to study the impact of additive noise on speech processing systems. It offers a diverse range of acoustic interference conditions. Four noise types are chosen: Babble, Factory, Pink, and White noise.

\subsection{Model and Training Configurations}
\subsubsection{Model Configuration}
The proposed NIL-TTS architecture adopts the architectural specifications of the official F5-TTS foundation model \cite{chen2025f5}, augmented with our environment-aware branch to facilitate dynamic vocal adaptation.

The Diffusion Transformer (DiT) backbone consists of $22$ layers with a hidden dimension of $d_{model}=1024$ and $16$ attention heads. The text encoder utilizes ConvNeXt V2 \cite{woo2023convnext} with a $512$-dimensional embedding space. For environmental sensing, we employ the official DAC encoder (44.1 kHz version) \cite{kumar2023high} as the noise encoder. This module processes $44.1$ kHz raw noise waveforms to extract $1024$-dimensional acoustic embeddings, which are then fused with a $128$-dimensional SNR-based MLP representation. 
The original ambient noise is resampled to $44.1$ kHz to align with the input requirements of the pre-trained DAC noise encoder.
Meanwhile, the core TTS pipeline and the vocoder operate consistently at $24$ kHz, and the model's internal representations map the conditions accordingly. The resulting environmental condition is injected into the hidden states of the DiT via the FiLM mechanism.
The model generates mel-spectrograms with $100$ frequency bins, which are subsequently converted into audio waveforms using a pre-trained Vocos vocoder \cite{siuzdak2024vocos}.

\begin{figure*}[!t]
\centering
\includegraphics[width=1.0\textwidth]{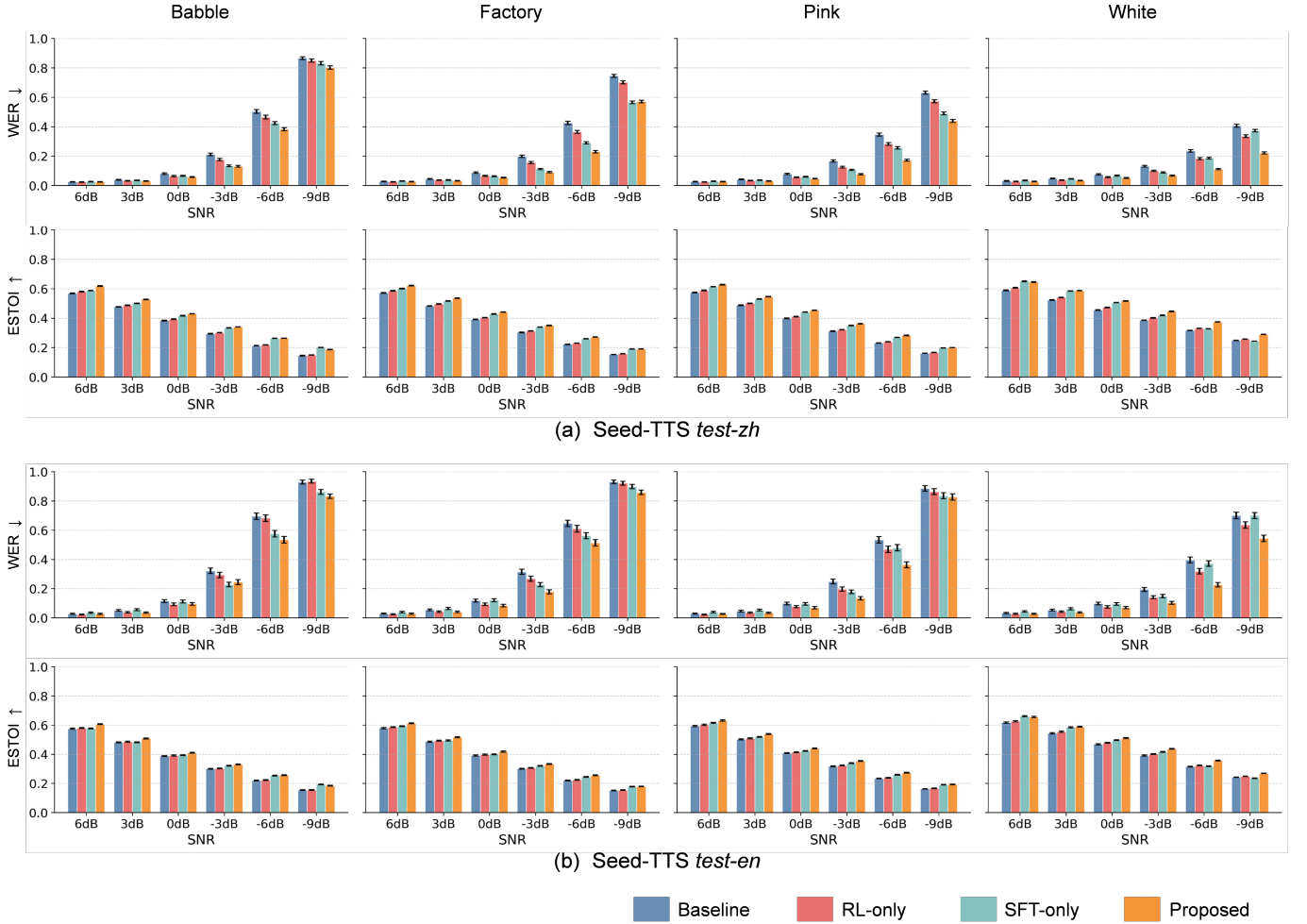}
\caption{Objective ASR transcription error rates (WER) and objective intelligibility metrics (ESTOI) across four noise types (Babble, Factory, Pink, and White) and six SNR levels (6 dB to -9 dB). Error bars denote 95\% confidence intervals. (a) Results from Seed-TTS \textit{test-zh}. (b) Results from Seed-TTS \textit{test-en}.}
\label{fig:object_result}
\end{figure*}

\subsubsection{Training Details}
\label{train_detail}
The training process is executed in two sequential stages: supervised fine-tuning (SFT) and on-policy reinforcement learning via Flow-GRPO.

During the SFT stage, we initialize the model with the weights of the pre-trained vanilla F5-TTS foundation model. To retain the high-fidelity phonetic and prosodic priors of the base model while focusing entirely on Lombard style injection, we freeze the DiT backbone and exclusively train the FiLM-based environment-aware modules. The model is optimized using the AdamW optimizer with a learning rate of $1 \times 10^{-4}$ and a sample-wise batch size of $8$. Training is conducted for $680,000$ steps on two NVIDIA A40 48GB GPUs. During this stage, the SNR is randomly sampled between -7.0 dB and 5.0 dB to establish the Lombard stylistic mapping.

In the post-training phase, the entire NIL-TTS architecture is unfrozen to allow for global strategy refinement. We reduce the learning rate to $1 \times 10^{-6}$ with a $400$-step linear warmup to stabilize policy updates. The entire NIL-TTS architecture is trained for 4,800 steps in total. For each prompt, the model generates a group of $G=24$ rollout trajectories using the SDE solver ($32$ NFE) with a diffusion coefficient of $a=0.4$ to encourage sufficient acoustic exploration. During post-training, the SNR is dynamically sampled from a uniform distribution $\mathcal{U}(-10, 15)$ dB.

The intelligibility reward $R_{acc}$ in Eq.~\ref{reward} is derived from the WER produced by the SenseVoice-Small ASR model \cite{an2024funaudiollm}, while the speaker similarity reward $R_{spk}$ is computed via cosine similarity using WeSpeaker embeddings \cite{wang2023wespeaker}. We set the reward weights to $\lambda_{acc}=6.0$ and $\lambda_{spk}=1.0$. The KL-divergence coefficient $\beta$ is set to $0.0$ in this study to facilitate aggressive policy exploration within the continuous acoustic space without being prematurely anchored by the SFT baseline. 

\subsection{Evaluation}
To quantitatively assess the performance of NIL-TTS, we conduct objective benchmarking using the Seed-TTS test set. We systematically perturb the synthesized speech samples with four representative noise typologies from the NoiseX-92 corpus \cite{varga1993assessment}: Babble, Factory, Pink, and White noise. To comprehensively evaluate the model's adaptive range, we simulate six distinct SNR regimes ranging from moderate interference to severe masking: {6, 3, 0, -3, -6, -9} dB. Following the synthesis of clean speech, the waveforms are mixed with the corresponding noise profiles to replicate situated acoustic environments.

For all evaluation tasks, we utilize a deterministic ODE solver with $32$ NFE. The Classifier-Free Guidance (CFG) strength is maintained at $2.0$. To optimize the generation trajectory, we apply Sway Sampling with a coefficient of $-1.0$ (following the default setup in F5-TTS \cite{chen2025f5}). Finally, the synthesized mel-spectrogram is vocoded at $24$ kHz.

\section{Results and Analysis}
In this section, we present the evaluation results of the proposed NIL-TTS framework, analyzing the intelligibility improvements and the underlying acoustic compensatory mechanisms discovered during the two-stage training paradigm.

\subsection{Intelligibility Improvements}

\subsubsection{Objective Intelligibility Evaluation} 
We assess objective intelligibility using two complementary metrics. 
All synthesized audio samples are resampled to 16 kHz prior to evaluation:

\begin{itemize}
\item \textbf{Word Error Rate (WER):} Derived from advanced ASR systems to reflect content preservation and noise robustness. We employ Paraformer-zh \cite{gao2023funasr} for Mandarin transcription and Whisper-large-v3 \cite{radford2023robust} for English.

\item \textbf{Extended Short-Time Objective Intelligibility (ESTOI) \cite{jensen2016algorithm}:} A signal-level metric that quantifies perceptual intelligibility under additive noise or processing distortions.
\end{itemize}

To demonstrate the extent to which our proposed model improves intelligibility, we use vanilla F5-TTS as the baseline and report results for the SFT-only model. Furthermore, to isolate the proposed noise-aware contribution from the already known benefit of RL post-training, we trained Flow-GRPO separately on the F5-TTS model (RL-only). This model also utilized WER and speaker-similarity rewards, but without noise-waveform and SNR conditioning or post-masker reward computation.

Fig.~\ref{fig:object_result} illustrates the performance of the baseline, RL-only, SFT-only, and the proposed NIL-TTS across diverse acoustic conditions. 
We establish statistical significance using paired utterance-level bootstrap resampling with 1,000 resamples, reporting the 95\% confidence intervals (CIs) across all conditions. The absolute intelligibility scores and corresponding CIs are detailed in the Appendix (Table~\ref{tab:detailed_scores}).

The results demonstrate that NIL-TTS achieves the highest objective intelligibility in nearly all evaluated conditions. Notably, at high SNRs (e.g., 6 dB), the improvement provided by the RL-enhanced model over the SFT model is relatively modest. This reflects a natural trade-off between intelligibility and naturalness: under mild noise conditions, excessive Lombard adaptation is unnecessary. However, the performance margin between NIL-TTS and the baseline widens significantly as the acoustic environment deteriorates.
This underscores the efficacy of our noise-induced style modification paradigm under extreme masking conditions. While all models exhibit performance degradation as SNR decreases, the Babble noise condition—characterized by non-stationary, speech-like masking—poses the greatest challenge. Conversely, white noise results in the smallest increase in WER because of its stationary characteristics.

NIL-TTS maintains intelligibility under severe noise. For instance, at a challenging $-6$ dB SNR on the Seed-TTS \textit{test-zh} dataset, the vanilla baseline degrades to a mean WER of $0.378$ (95\% CI: $[0.372, 0.383]$). The SFT-only model improves this to $0.289$ (95\% CI: $[0.285, 0.294]$), whereas the proposed NIL-TTS significantly reduces the WER to $0.224$ (95\% CI: $[0.219, 0.228]$). Similar statistical superiority is observed in the ESTOI metric, where NIL-TTS achieves $0.298$ (95\% CI: $[0.297, 0.299]$) at $-6$ dB SNR, outperforming the SFT-only model ($0.279$, 95\% CI: $[0.278, 0.280]$) and the baseline ($0.246$, 95\% CI: $[0.244, 0.247]$).

\begin{DIFnomarkup}
\begin{table*}[!t]
\caption{Intelligibility performance in terms of WER and ESTOI, averaged across four noise types at SNRs of 0, \(-3\), and \(-6\) dB on the Seed-TTS test set. {\dag\enspace}denotes models implemented by the authors.}
\label{tab:compare-other}
\centering
\small
\renewcommand{\arraystretch}{1.25}   

\begin{tabular}{l *{12}{c}}   
\toprule
\multirow{4}{*}{\textbf{Conditions}} 
& \multicolumn{6}{c}{\textbf{Seed-TTS \textit{test-zh}}} & \multicolumn{6}{c}{\textbf{Seed-TTS \textit{test-en}}} \\
\cmidrule(lr){2-7}  \cmidrule(lr){8-13} 
& \multicolumn{3}{c}{\textbf{WER} $\downarrow$} & \multicolumn{3}{c}{\textbf{ESTOI} $\uparrow$} & 
\multicolumn{3}{c}{\textbf{WER} $\downarrow$} & \multicolumn{3}{c}{\textbf{ESTOI} $\uparrow$}\\
\cmidrule(lr){2-4}  \cmidrule(lr){5-7} \cmidrule(lr){8-10}  \cmidrule(lr){11-13} 
& 0~dB & -3~dB & -6~dB & 0~dB & -3~dB & -6~dB & 0~dB & -3~dB & -6~dB & 0~dB & -3~dB & -6~dB\\
\midrule
F5-TTS \cite{chen2025f5} & 0.081 & 0.177 & 0.378 & 0.407 & 0.324 & 0.246 & 0.107 & 0.270 & 0.568 & 0.414 & 0.327 & 0.247\\

\addlinespace[2pt]
\hdashline[2pt/2pt] 
\addlinespace[2pt]

F5-TTS + SS & 0.057 & 0.123 & 0.278 & 0.420 & 0.339 & 0.264 & 0.086 & 0.183 & 0.498 & 0.394 & 0.321 & 0.254\\
F5-TTS + SSDRC & 0.058 & 0.120 & 0.263 & 0.448 & 0.347 & 0.270 & 0.089 & 0.176 & 0.484 & 0.413 & 0.333 & 0.257\\

\addlinespace[2pt]
\hdashline[2pt/2pt] 
\addlinespace[2pt]

MetaVoice-1B & - & - & - & - & - & - & 0.419 & 0.651 & 0.895 & 0.420 & 0.331 & 0.248\\
MetaVoice-1B Lomb\dag \cite{lobato2025gradual} & - & - & - & - & - & - & 0.420 & 0.640 & 0.873 & 0.428 & 0.337 & 0.253\\
F5-TTS-Style Base\dag \cite{akti2026lombard} & 0.092 & 0.190 & 0.386 & 0.419 & 0.337 & 0.258 & 0.181 & 0.364 & 0.654 & 0.425 & 0.339 & 0.258\\
F5-TTS-Style Lomb\dag \cite{akti2026lombard} & 0.099 & 0.199 & 0.396 & 0.417 & 0.335 & 0.257 & 0.200 & 0.392 & 0.678 & 0.423 & 0.337 & 0.256\\

\addlinespace[2pt]
\hdashline[2pt/2pt] 
\addlinespace[2pt]

RL-only & 0.061 & 0.140 & 0.323 & 0.420 & 0.334 & 0.254 & 0.084 & 0.225 & 0.519 & 0.420 & 0.333 & 0.252\\

SFT-only & 0.064 & 0.110 & 0.289 & 0.448 & 0.361 & 0.279 & {0.105} & {0.195} & {0.497} & {0.428} & {0.349} & {0.269}\\

NIL-TTS & \textbf{0.052} & \textbf{0.091} & \textbf{0.224} & \textbf{0.460} & \textbf{0.375} & \textbf{0.298} & \textbf{0.079} & \textbf{0.165} & \textbf{0.434} & \textbf{0.445} & \textbf{0.364} & \textbf{0.285}\\

\bottomrule
\end{tabular}
\end{table*}
\end{DIFnomarkup}

The addition of the \textit{RL-only} baseline isolates the contribution of the proposed environmental conditioning mechanism. While applying Flow-GRPO with WER and speaker-similarity rewards (without noise conditioning) does improve intelligibility over the vanilla baseline, it falls short of the NIL-TTS performance. This confirms that part of the intelligibility gain in adverse conditions is derived from the explicit noise-awareness, rather than the RL post-training alone.

Notably, although the Flow-GRPO stage is conducted exclusively on Mandarin data, we observe consistent intelligibility gains on zero-shot English synthesis under adverse noise conditions.
This empirical transfer suggests that the learned acoustic compensation patterns are not strictly tied to language-specific phonetic content. The observed cross-lingual improvements indicate that the RL-driven adaptation primarily operates in acoustic space rather than lexical space.

Furthermore, we compared the objective intelligibility of our proposed NIL-TTS against signal-processing-based modifications and established Lombard speech synthesis methods. 

Two signal-processing-based modifications were utilized:

\begin{itemize}
    \item \textbf{SS:} The baseline is processed using adaptive spectral shaping (SS) \cite{zorila2012speech}, which artificially enhances formants and flattens spectral tilt.
    \item \textbf{SSDRC:} The baseline is processed using both spectral shaping and dynamic range compression (SSDRC) \cite{zorila2012speech}, which reallocates signal energy across both frequency and time domains.
\end{itemize}

We selected two models capable of zero-shot Lombard speech synthesis:

\begin{itemize}
    \item \textbf{MetaVoice-1B Lomb \cite{lobato2025gradual}:} Built upon the MetaVoice-1B\footnote{\url{https://github.com/metavoiceio/metavoice-src}} foundation model, this approach learns a transformation between plain speech embeddings and their Lombard variants. Following the original methodology, we trained a feedforward speaker transformation model using paired data from the LombardGrid corpus to map plain speaker embeddings to Lombard embeddings for zero-shot synthesis. As this model only supports English speech synthesis, only the \textit{test-en} results are reported.
    \item \textbf{F5-TTS-Style \cite{akti2026lombard}:} This method, also based on the F5-TTS architecture, introduces an additional style encoder to model speaker and style features via FiLM injection. Instead of relying on explicit Lombard training data, it controls the Lombard effect by manipulating relevant principal components (via PCA mapping) in the style embedding space. Following the authors' protocol, we fine-tuned F5-TTS with the style encoder on the Emilia dataset \cite{11175027} (evaluated as F5-TTS-Style Base). We then synthesized Lombard speech by shifting the relevant PCA components to the "Very Loud" level specified in the original paper (evaluated as F5-TTS-Style Lomb).
\end{itemize}

As shown in Table~\ref{tab:compare-other}, both SS and SSDRC effectively improve speech intelligibility over the baseline. This validates the utility of traditional acoustic enhancements such as energy reallocation. The MetaVoice-1B Lomb provides marginal improvements in severe noise. In contrast, the F5-TTS-Style framework degrades the WER compared to its base version (e.g., increasing WER from $0.092$ to $0.099$ at 0 dB SNR on \textit{test-zh}). The proposed NIL-TTS consistently outperforms both the signal-processing techniques and the neural Lombard baselines across all evaluated SNRs in both the Mandarin and English subsets.

The performance gap stems from the distinct mechanisms underlying these approaches. While SS and SSDRC apply deterministic acoustic filters uniformly across the generated audio, they lack contextual and phonetic awareness. 
Similarly, existing neural approaches such as F5-TTS-Style simulate the Lombard effect by shifting principal components associated with loudness and clarity. This method manipulates embeddings, resulting in a coarse simulation of the Lombard style that lacks a direct optimization objective for speech intelligibility. Consequently, the structural modifications to the speech signal may actually degrade recognition accuracy in severe noise compared to its base version.

Rather than applying a rigid post-hoc modification or a static style vector, our model leverages its environment-aware sensing pathway to autonomously shift vocal energy to frequency bands less affected by the specific ambient noise profile. Furthermore, the superiority of the full NIL-TTS over the \textit{RL-only} and \textit{SFT-only} configurations confirms that our two-stage training paradigm is essential for yielding these natural and robust intelligibility gains.

\subsubsection{Subjective Intelligibility Evaluation}
We conducted a blind subjective intelligibility test following the ITU-T P.807 standard \cite{rec2016p}. We focused on Babble noise due to its representative complexity in objective tests. Three SNR levels (0~dB, -3~dB, and -6~dB) were selected, representing a reasonable range that causes significant fluctuations in speech intelligibility. Our objective evaluations comprehensively demonstrated the superiority of NIL-TTS over traditional signal-processing methods and other zero-shot synthetic models. Therefore, the subjective study specifically focused on isolating the perceptual contributions of our architectural components. 

\begin{figure}[!t]
\centering
\includegraphics[width=0.4\textwidth]{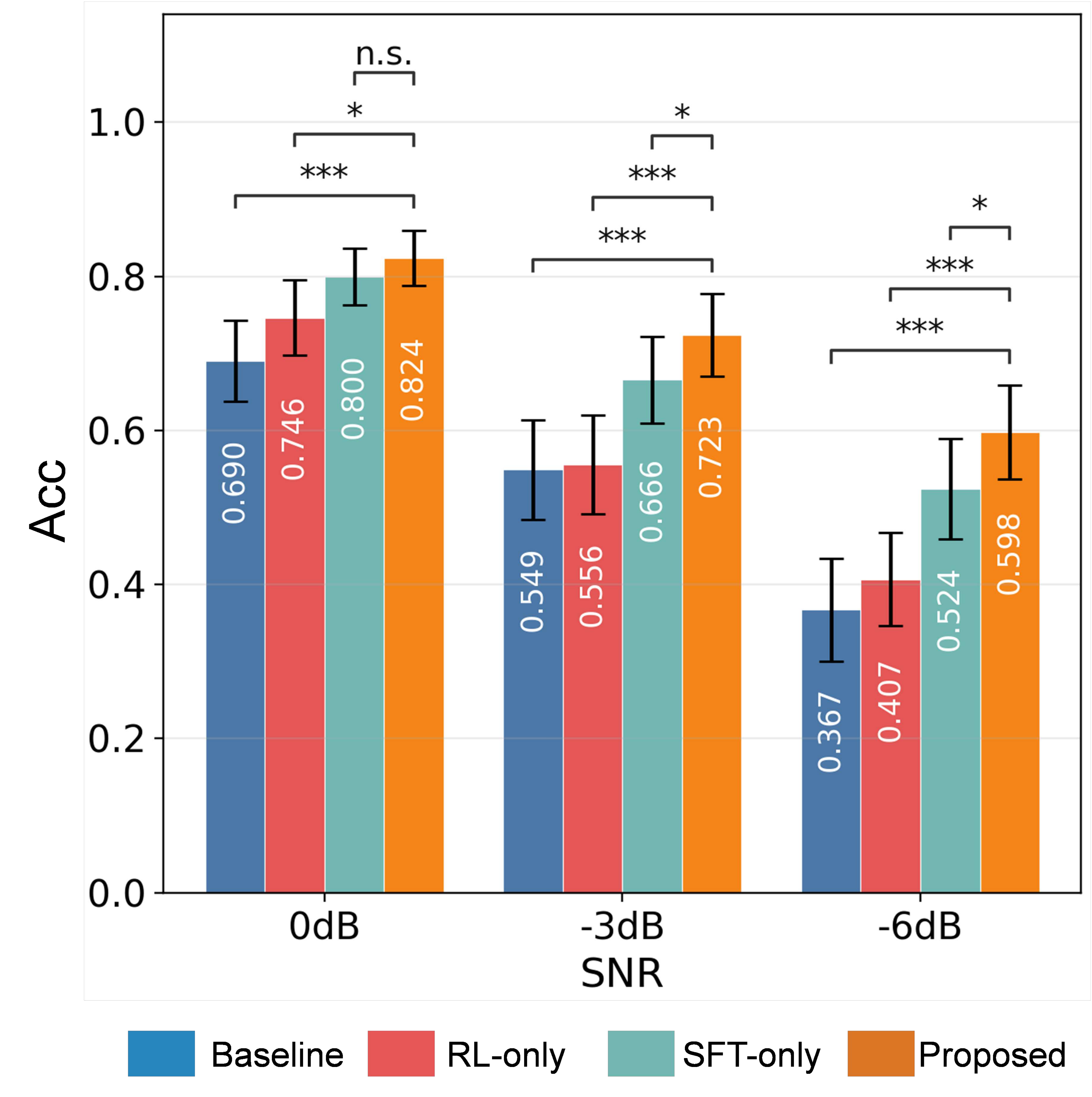}
\caption{Subjective recognition accuracy across 3 SNR levels (0 dB, -3 dB, and -6 dB) under Babble noise. Error bars denote 95\% confidence intervals.}
\label{fig:subject_result}
\end{figure}

Twelve listeners were tasked with transcribing 360 randomly selected Mandarin speech clips, balanced across four model configurations (baseline, RL-only, SFT-only, and the proposed NIL-TTS) and three SNRs. Each participant's test set was carefully designed to avoid memory effects resulting from repeated sentence content, which may introduce bias. To ensure a fair comparison, the text content used for synthesis is identical across all evaluated models. 

As shown in Fig.~\ref{fig:subject_result}, the paired $t$-test results indicate that the proposed NIL-TTS yields statistically significant subjective intelligibility improvements over the vanilla baseline across all evaluated SNRs ($p < 0.001$ for all). Crucially, the inclusion of the ablation models isolates and clarifies the source of the perceptual gain. While the RL-only model provides a marginal improvement over the baseline, the proposed model significantly outperforms it at all SNRs ($p=0.018$ at 0~dB; $p < 0.001$ at -3~dB and -6~dB), confirming that RL-based post-training alone is insufficient without our explicit environmental conditioning mechanism. Furthermore, when compared to the supervised Lombard adaptation (SFT-only), the proposed framework shows no statistically significant difference under mild noise (0~dB, $p=0.368$), but it achieves significant perceptual improvements under more severe masking conditions: -3~dB ($p=0.012$) and -6~dB ($p=0.027$).

The absolute transcription accuracy gain provided by the proposed model over the baseline increases monotonically as the SNR drops, reaching a substantial peak at -6~dB (improving from $0.367$ to $0.598$). These results indicate that the improvements observed in objective metrics are consistently reflected in subjective listening tests, successfully verifying that the full environment-aware conditioning and two-stage training mechanism of NIL-TTS are necessary for maximizing perceptible human intelligibility under adverse conditions.

\begin{figure*}[!t]
\centering
\includegraphics[width=1.0\textwidth]{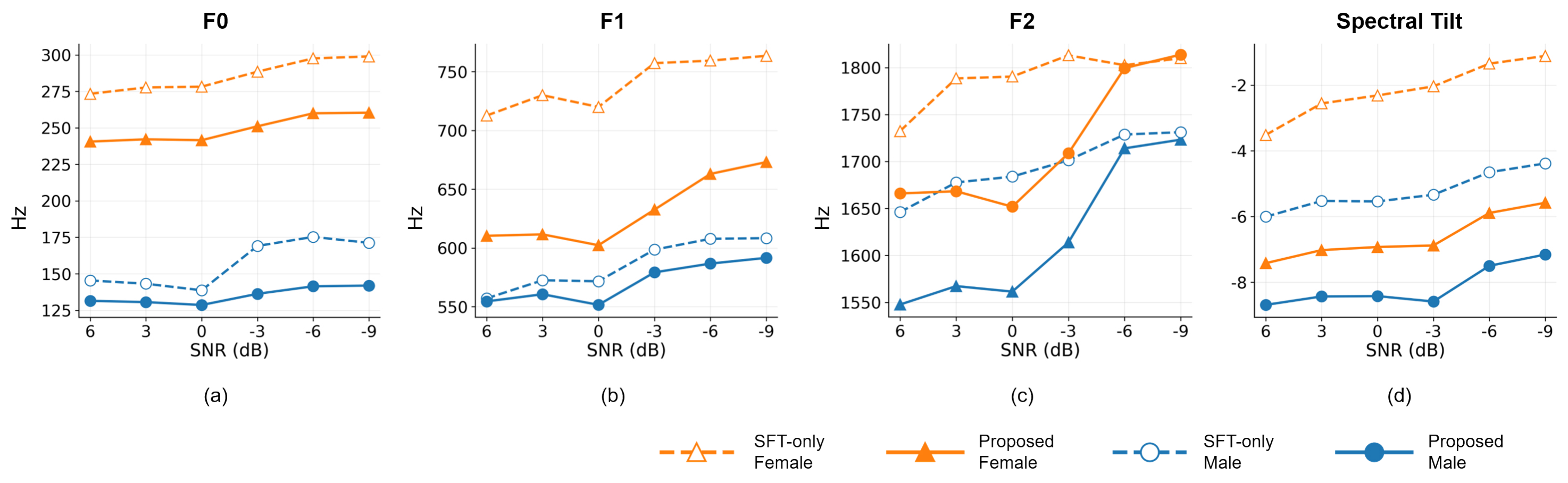}
\caption{Comparison of acoustic features produced by the SFT-only and proposed NIL-TTS models across different SNR levels for female and male speakers. The panels show (a) fundamental frequency $F_0$, (b) first formant $F_1$, (c) second formant $F_2$, and (d) spectral tilt. Dashed lines with open markers denote the SFT-only model, whereas solid lines with filled markers denote the proposed model after Flow-GRPO post-training.}
\label{fig:acoustic_metrics}
\end{figure*}

\begin{figure*}[!t]
\centering
\includegraphics[width=1.0\textwidth]{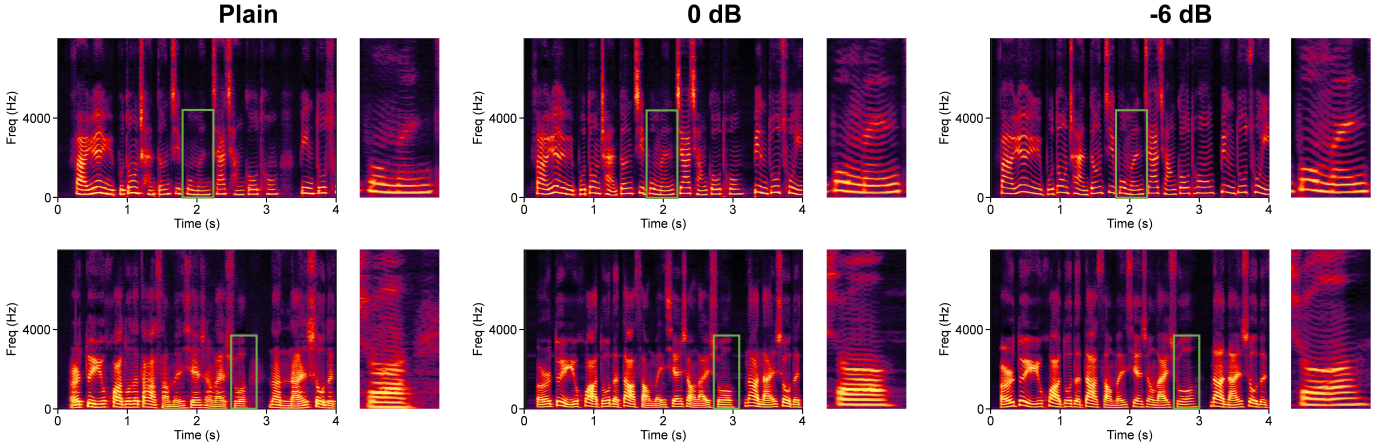}
\caption{Spectrogram comparison illustrating emergent Lombard characteristics. All samples are synthesized by the proposed NIL-TTS. Each row corresponds to a distinct test utterance generated using the same text and speaker reference, but conditioned on varying ambient noise levels (Plain, 0 dB, and -6 dB). The \textit{Plain} condition represents standard, non-adapted speech generated by disabling the FiLM branch. Under noise, the model exhibits upward harmonic shifts and pronounced energy concentration in higher frequency bands.}
\label{fig:spectrograms}
\end{figure*}

\subsection{Acoustic Parameter Analysis}
To examine how Flow-GRPO modifies the acoustic adaptation learned during supervised fine-tuning, we compare the SFT-only model with the complete NIL-TTS model after Flow-GRPO post-training across varying noise intensities. We evaluate four primary acoustic correlates of the Lombard effect: fundamental frequency ($F_0$), the first and second formants ($F_1$ and $F_2$), and spectral tilt. The fundamental frequency $F_0$, reflecting vocal fold vibration rate, is estimated using the pYIN algorithm \cite{mauch2014pyin}. The resonance properties of the vocal tract, represented by formants $F_1$ and $F_2$, are extracted via Linear Predictive Coding (LPC) \cite{snell1993formant}. Spectral tilt, which indicates the distribution of energy across the frequency spectrum, is quantified as the slope of a linear regression line fitted to the Long-Term Average Spectrum (LTAS) \cite{nordenberg2004effect} on a logarithmic frequency scale.

As shown in Fig.~\ref{fig:acoustic_metrics}, both models exhibit systematic SNR-dependent acoustic changes, although their absolute parameter values and adaptation patterns differ. For both female and male speakers, $F_0$, $F_1$, and $F_2$ generally increase as the SNR decreases. The SFT-only model produces higher absolute $F_0$, $F_1$, and $F_2$ values, whereas the proposed model retains lower absolute values while preserving systematic SNR-dependent increases.
Spectral tilt also becomes progressively less negative for both models as the SNR decreases, indicating a flatter spectral envelope and relatively greater high-frequency energy. However, the proposed model retains a steeper spectral tilt than the SFT-only model across the evaluated conditions.

These results indicate that Flow-GRPO does not simply amplify every Lombard-related acoustic cue acquired during supervised fine-tuning. Instead, it rebalances the acoustic adaptation strategy, moderating the changes in $F_0$, $F_1$, $F_2$, and spectral tilt. Together with the objective and subjective intelligibility improvements reported above, this comparison suggests that the reinforcement learning stage discovers a more selective combination of acoustic adjustments for improving robustness in noise.


Spectrogram analysis (Fig.~\ref{fig:spectrograms}) visually confirms these Lombard characteristics. As the SNR decreases, the harmonic components shift upward, and the energy becomes more concentrated at higher frequencies. This observation is consistent with the within-model reduction in spectral tilt under increasingly severe noise, which indicates a redistribution of speech energy toward higher-frequency regions. Together, the acoustic-feature and spectrogram analyses suggest that Flow-GRPO refines the relative contributions of pitch, formant, and spectral cues rather than uniformly increasing their magnitudes.

\subsection{Computational Efficiency Analysis}
We examine the computational overhead of NIL-TTS relative to the baselines. All F5-TTS variants use deterministic ODE sampling with 32 function evaluations and a classifier-free guidance (CFG) scale of 2.0.

The vanilla F5-TTS baseline operates with 336M parameters and achieves an efficient RTF of 0.26. The integration of our environmental pathway introduces a modest overhead of parameters, bringing the total to 382M. Despite this addition, the inference speed is essentially unaffected, maintaining a highly competitive RTF of 0.27. 

MetaVoice-1B Lomb has 1.2B parameters and an RTF of 3.04, making it unsuitable for low-latency conversational applications. Although F5-TTS-Style Lomb has a computational footprint similar to that of NIL-TTS (387M parameters, RTF 0.27), it lacks the explicit intelligibility optimization provided by our framework.

Ultimately, these results confirm that our proposed environmental conditioning and two-stage training paradigm impose a negligible inference overhead (an RTF increase of merely 0.01 over the vanilla baseline), ensuring that NIL-TTS remains highly practical for real-world, real-time deployment.

\begin{table}[!t]
\caption{Comparison of computational costs. \#Param. denotes the number of learnable parameters. The real-time factor (RTF) is measured for generating 10-s utterances on an NVIDIA RTX 3090 GPU.}
\label{tab:computational cost}
\centering
\small
\renewcommand{\arraystretch}{1.25}   

\begin{tabular}{lrr}   
\toprule
\textbf{Config} & \textbf{\#Param}. & \textbf{RTF} $\downarrow$ \\
\midrule

F5-TTS & 336M & 0.26 \\
\addlinespace[2pt]
\hdashline[2pt/2pt] 
\addlinespace[2pt]
MetaVoice-1B Lomb & 1.2B & 3.04 \\
F5-TTS-Style Lomb & 387M & 0.27 \\
NIL-TTS & 382M & 0.27 \\

\bottomrule
\end{tabular}
\end{table}

\begin{DIFnomarkup}
\begin{table}[!t]
\caption{WER obtained using different Flow-GRPO hyperparameter settings, averaged across four noise types at SNRs of 0, \(-3\), and \(-6\) dB on the Seed-TTS test set.}
\label{tab:ablation}
\centering
\small
\renewcommand{\arraystretch}{1.25}   

\begin{tabular}{l *{6}{c}}   
\toprule
\multirow{2.5}{*}{\textbf{Config}} 
& \multicolumn{3}{c}{\textbf{Seed-TTS \textit{test-zh}}} & \multicolumn{3}{c}{\textbf{Seed-TTS \textit{test-en}}} \\
\cmidrule(lr){2-4}  \cmidrule(lr){5-7} 
& 0~dB & -3~dB & -6~dB & 0~dB & -3~dB & -6~dB \\
\midrule

NIL-TTS & \textbf{0.052} & 0.091 & \textbf{0.224} & \textbf{0.079} & 0.165 & \textbf{0.434} \\

\addlinespace[2pt]
\hdashline[2pt/2pt] 
\multicolumn{7}{c}{\textbf{KL penalty}} \\   
\hdashline[2pt/2pt]    
\addlinespace[2pt]
$\beta$=0.01 & 0.063 & \textbf{0.088} & 0.235 & 0.083 & 0.214 & 0.460 \\
$\beta$=0.1  & 0.056 & 0.102 & 0.252 & 0.100 & \textbf{0.158} & 0.435 \\

\addlinespace[2pt]
\hdashline[2pt/2pt] 
\multicolumn{7}{c}{\textbf{NFE steps}} \\   
\hdashline[2pt/2pt]  
\addlinespace[2pt]
NFE=8 & 0.065 & 0.130 & 0.292 & 0.098 & 0.226 & 0.506 \\
NFE=16 & 0.058 & 0.106 & 0.247 & 0.102 & 0.204 & 0.485 \\

\addlinespace[2pt]
\hdashline[2pt/2pt] 
\multicolumn{7}{c}{\textbf{Noise Level \textit{a}}} \\   
\hdashline[2pt/2pt]    
\addlinespace[2pt]
\textit{a}=0.2 & 0.058 & 0.111 & 0.259 & 0.097 & 0.202 & 0.482 \\
\textit{a}=0.6 & 0.062 & 0.103 & 0.260 & 0.096 & 0.194 & 0.501 \\

\bottomrule
\end{tabular}
\end{table}
\end{DIFnomarkup}

\subsection{Ablation Studies on Post-Training Design Choices}
The sparsity of rewards in reinforcement learning can lead to training instability. To evaluate the robustness of our post-training phase, we systematically ablated key Flow-GRPO hyperparameters, analyzing their impact on final intelligibility (Table~\ref{tab:ablation}).

First, we examined the KL divergence penalty ($\beta$). While $\beta=0.01$ and $\beta=0.1$ achieve marginally lower WERs at specific SNR levels, $\beta=0.0$ yields superior overall intelligibility. We empirically found that introducing a non-zero KL penalty overly restricts the policy update in the continuous acoustic space, limiting the model’s ability to discover effective compensatory strategies. Therefore, we adopt $\beta$=0 and rely on structural regularization (SFT initialization and speaker similarity constraints) to maintain stability.
In contrast to discrete autoregressive policies, the flow-based generative model operates in a continuous trajectory space governed by diffusion priors. The stochastic SDE sampling implicitly constrains exploration around the learned probability path, reducing the risk of catastrophic drift. This allows unconstrained policy optimization without requiring explicit KL regularization.

Regarding the Number of Function Evaluations (NFE) during the SDE rollout, our experiments demonstrate that utilizing $32$ steps (matching the inference configuration) yields better WER metrics than $8$ or $16$ steps, though the experiments in the original Flow-GRPO study \cite{liu2026flow} suggest that the usage of fewer rollout steps results in faster training convergence and better metrics. This discrepancy between our experiments and the original Flow-GRPO may stem from differences in the model’s task and the nature of the metrics. The original Flow-GRPO is designed to make generated images better align with human preferences. In contrast, our work requires generating clearer, more perceptible acoustic cues, which may rely on a finer-grained sampling process.

Finally, we evaluated the diffusion coefficient $a$, which dictates the exploration intensity of the Wiener process during SDE sampling. An intermediate value of $a=0.4$ outperforms both $a=0.2$ and $a=0.6$. This indicates that moderate stochasticity is critical: excessive noise ($a=0.6$) disrupts the phonetic structure required by the ASR reward, while insufficient noise ($a=0.2$) fails to provide the policy with the diverse vocal trajectories needed to discover effective Lombard compensations.

\begin{table*}[!t]
\caption{Recognition performance (WER) of modern ASR models on plain (P) and Lombard speech (L1 and L2) under diverse acoustic conditions.}
\label{tab:asr_validation}
\centering
\small
\renewcommand{\arraystretch}{1.0}   

\begin{tabular}{l c *{12}{r}}  
\toprule
\multirow{2.5}{*}{\textbf{ASR Model}} & \multirow{2.5}{*}{\textbf{SNR (dB)}} 
& \multicolumn{3}{c}{\textbf{White}} 
& \multicolumn{3}{c}{\textbf{Pink}} 
& \multicolumn{3}{c}{\textbf{Babble}} 
& \multicolumn{3}{c}{\textbf{Factory}} \\
\cmidrule(lr){3-5} \cmidrule(lr){6-8} \cmidrule(lr){9-11} \cmidrule(lr){12-14}
& & P & L1 & L2 
& P & L1 & L2 
& P & L1 & L2 
& P & L1 & L2 \\
\midrule

\multirow{4}{*}{SenseVoice-Small}
& 5  & 0.08 & 0.08 & \textbf{0.07} & 0.06 & 0.07 & 0.06 & 0.06 & 0.06 & 0.06 & 0.08 & 0.08 & 0.08 \\
& 0  & 0.15 & 0.14 & \textbf{0.12} & 0.16 & 0.15 & \textbf{0.13} & 0.18 & 0.17 & \textbf{0.15} & 0.22 & 0.21 & \textbf{0.17} \\
& -3 & 0.23 & 0.22 & \textbf{0.19} & 0.30 & 0.29 & \textbf{0.24} & 0.40 & 0.38 & \textbf{0.30} & 0.43 & 0.42 & \textbf{0.32} \\
& -5 & 0.31 & 0.30 & \textbf{0.26} & 0.47 & 0.46 & \textbf{0.37} & 0.65 & 0.61 & \textbf{0.46} & 0.67 & 0.65 & \textbf{0.51} \\

\addlinespace[6pt]

\multirow{4}{*}{Paraformer}
& 5  & 0.09 & 0.09 & 0.09 & 0.07 & 0.07 & 0.07 & 0.06 & 0.06 & 0.06 & 0.09 & 0.09 & 0.09 \\
& 0  & 0.16 & 0.15 & \textbf{0.13} & 0.16 & 0.15 & \textbf{0.13} & 0.17 & 0.16 & \textbf{0.14} & 0.20 & 0.19 & \textbf{0.17} \\
& -3 & 0.23 & 0.22 & \textbf{0.19} & 0.29 & 0.28 & \textbf{0.23} & 0.35 & 0.33 & \textbf{0.27} & 0.37 & 0.35 & \textbf{0.29} \\
& -5 & 0.30 & 0.29 & \textbf{0.25} & 0.42 & 0.41 & \textbf{0.34} & 0.53 & 0.51 & \textbf{0.41} & 0.53 & 0.51 & \textbf{0.42} \\

\addlinespace[6pt]

\multirow{4}{*}{FireRedASR-AED}
& 5  & 0.08 & 0.08 & 0.08 & 0.07 & 0.07 & 0.07 & 0.06 & 0.06 & 0.06 & 0.08 & 0.08 & 0.08 \\
& 0  & 0.13 & 0.13 & \textbf{0.11} & 0.14 & 0.13 & \textbf{0.11} & 0.14 & 0.13 & \textbf{0.11} & 0.18 & 0.17 & \textbf{0.15} \\
& -3 & 0.20 & 0.19 & \textbf{0.16} & 0.26 & 0.25 & \textbf{0.19} & 0.30 & 0.28 & \textbf{0.21} & 0.34 & 0.32 & \textbf{0.26} \\
& -5 & 0.27 & 0.26 & \textbf{0.21} & 0.41 & 0.39 & \textbf{0.30} & 0.49 & 0.46 & \textbf{0.33} & 0.52 & 0.50 & \textbf{0.39} \\

\addlinespace[6pt]

\multirow{4}{*}{Whisper}
& 5  & 0.18 & 0.17 & \textbf{0.15} & 0.15 & 0.15 & \textbf{0.14} & 0.12 & 0.13 & 0.12 & 0.18 & 0.17 & \textbf{0.15} \\
& 0  & 0.29 & 0.31 & \textbf{0.25} & 0.31 & 0.31 & \textbf{0.27} & 0.32 & 0.31 & \textbf{0.25} & 0.38 & 0.40 & \textbf{0.32} \\
& -3 & 0.43 & 0.41 & \textbf{0.39} & 0.50 & 0.49 & \textbf{0.45} & 0.56 & 0.53 & \textbf{0.46} & 0.64 & 0.62 & \textbf{0.51} \\
& -5 & 0.54 & 0.52 & \textbf{0.50} & 0.68 & 0.65 & \textbf{0.62} & 0.81 & 0.76 & \textbf{0.65} & 0.80 & 0.78 & \textbf{0.71} \\

\bottomrule
\end{tabular}
\end{table*}

\section{Conclusion}
In this work, we introduced NIL-TTS, a noise-aware Lombard speech synthesis framework that integrates FiLM-based environmental conditioning into a flow-matching backbone to enable autonomous vocal adaptation. Through a two-stage training paradigm consisting of supervised fine-tuning followed by Flow-GRPO post-training, the model successfully optimizes for intelligibility and speaker consistency beyond simple Lombard stylistic imitation. Experimental results on the Seed-TTS benchmark confirm that NIL-TTS significantly enhances speech intelligibility in adverse conditions. Moreover, the framework recovers key acoustic correlates of the human Lombard effect, providing a practical framework for noise-aware speech synthesis in challenging real-world environments.


\begin{table*}[htbp]
\centering
\small
\caption{Absolute WER and ESTOI scores with 95\% confidence intervals obtained through paired utterance-level bootstrap resampling under different SNR conditions.}
\label{tab:all-results}
\renewcommand{\arraystretch}{1.15} 

\begin{tabular}{llcccc}
\toprule
\textbf{Metric} & \textbf{SNR} & \textbf{Baseline} & \textbf{RL-only} & \textbf{SFT-only} & \textbf{Proposed} \\
\midrule
\multicolumn{6}{c}{\textbf{Seed-TTS \textit{test-zh}}} \\
\midrule
\multirow{6}{*}{\textbf{WER} $\downarrow$} 
 & 6dB  & 0.027 [0.026, 0.028] & \textbf{0.025} [0.024, 0.026] & 0.030 [0.029, 0.031] & 0.026 [0.025, 0.027] \\
 & 3dB  & 0.043 [0.041, 0.044] & 0.035 [0.033, 0.036] & 0.038 [0.037, 0.040] & \textbf{0.032} [0.031, 0.033] \\
 & 0dB  & 0.081 [0.078, 0.083] & 0.061 [0.059, 0.062] & 0.064 [0.062, 0.066] & \textbf{0.052} [0.051, 0.054] \\
 & -3dB & 0.177 [0.173, 0.181] & 0.140 [0.136, 0.143] & 0.110 [0.108, 0.113] & \textbf{0.091} [0.089, 0.093] \\
 & -6dB & 0.378 [0.372, 0.383] & 0.323 [0.318, 0.329] & 0.289 [0.285, 0.294] & \textbf{0.224} [0.219, 0.228] \\
 & -9dB & 0.662 [0.656, 0.669] & 0.615 [0.608, 0.621] & 0.565 [0.558, 0.571] & \textbf{0.508} [0.501, 0.515] \\
\cmidrule{1-6}
\multirow{6}{*}{\textbf{ESTOI} $\uparrow$} 
 & 6dB  & 0.576 [0.575, 0.577] & 0.590 [0.588, 0.591] & 0.613 [0.612, 0.614] & \textbf{0.628} [0.627, 0.629] \\
 & 3dB  & 0.493 [0.491, 0.494] & 0.506 [0.505, 0.507] & 0.533 [0.531, 0.534] & \textbf{0.549} [0.548, 0.550] \\
 & 0dB  & 0.407 [0.406, 0.408] & 0.420 [0.418, 0.421] & 0.448 [0.447, 0.449] & \textbf{0.460} [0.459, 0.462] \\
 & -3dB & 0.324 [0.323, 0.325] & 0.334 [0.333, 0.336] & 0.361 [0.359, 0.362] & \textbf{0.375} [0.374, 0.376] \\
 & -6dB & 0.246 [0.244, 0.247] & 0.254 [0.253, 0.255] & 0.279 [0.278, 0.280] & \textbf{0.298} [0.297, 0.299] \\
 & -9dB & 0.177 [0.176, 0.178] & 0.183 [0.182, 0.184] & 0.208 [0.207, 0.209] & \textbf{0.216} [0.215, 0.218] \\

\midrule
\multicolumn{6}{c}{\textbf{Seed-TTS \textit{test-en}}} \\
\midrule
\multirow{6}{*}{\textbf{WER} $\downarrow$} 
 & 6dB  & 0.030 [0.028, 0.032] & \textbf{0.024} [0.022, 0.026] & 0.039 [0.037, 0.042] & 0.027 [0.025, 0.029] \\
 & 3dB  & 0.051 [0.048, 0.054] & 0.040 [0.037, 0.042] & 0.058 [0.055, 0.062] & \textbf{0.037} [0.035, 0.039] \\
 & 0dB  & 0.107 [0.102, 0.112] & 0.084 [0.080, 0.088] & 0.105 [0.100, 0.110] & \textbf{0.079} [0.075, 0.083] \\
 & -3dB & 0.270 [0.260, 0.278] & 0.225 [0.216, 0.232] & 0.195 [0.188, 0.202] & \textbf{0.165} [0.158, 0.172] \\
 & -6dB & 0.568 [0.556, 0.579] & 0.519 [0.508, 0.531] & 0.497 [0.487, 0.508] & \textbf{0.434} [0.423, 0.445] \\
 & -9dB & 0.861 [0.853, 0.871] & 0.839 [0.829, 0.848] & 0.824 [0.814, 0.833] & \textbf{0.765} [0.755, 0.775] \\
\cmidrule{1-6}
\multirow{6}{*}{\textbf{ESTOI} $\uparrow$} 
 & 6dB  & 0.591 [0.589, 0.593] & 0.598 [0.596, 0.601] & 0.611 [0.610, 0.613] & \textbf{0.626} [0.624, 0.628] \\
 & 3dB  & 0.503 [0.501, 0.505] & 0.510 [0.508, 0.512] & 0.520 [0.518, 0.522] & \textbf{0.538} [0.536, 0.540] \\
 & 0dB  & 0.414 [0.412, 0.416] & 0.420 [0.418, 0.423] & 0.428 [0.426, 0.430] & \textbf{0.445} [0.443, 0.447] \\
 & -3dB & 0.327 [0.325, 0.329] & 0.333 [0.331, 0.335] & 0.349 [0.347, 0.351] & \textbf{0.364} [0.362, 0.366] \\
 & -6dB & 0.247 [0.245, 0.249] & 0.252 [0.250, 0.254] & 0.269 [0.268, 0.271] & \textbf{0.285} [0.284, 0.287] \\
 & -9dB & 0.177 [0.175, 0.179] & 0.181 [0.179, 0.182] & 0.199 [0.198, 0.200] & \textbf{0.206} [0.205, 0.208] \\
\bottomrule
\label{tab:detailed_scores}
\end{tabular}
\end{table*}

{\appendices
\section{Validation of ASR-Based Rewards for Inducing Lombard Speech under Adverse Conditions}
A primary theoretical concern in reinforcement learning-based TTS is whether the objective feedback provided by ASR models is consistent with human subjective intelligibility, particularly regarding the Lombard effect. While biological Lombard adaptations are known to significantly enhance perceptual clarity for human listeners, it remains to be verified whether modern ASR architectures exhibit a similar sensitivity. If ASR models fail to achieve higher recognition accuracy for Lombard speech compared to plain speech in noisy environments, utilizing ASR-based rewards might lead to misaligned policies that do not truly enhance intelligibility. 

To address this, we conducted a feasibility study to verify the "reward-signal validity" of the ASR-based feedback mechanism. We utilized the EMALG corpus, which provides recordings of plain (P, 40~dB), moderate Lombard (L1, 55~dB), and strong Lombard (L2, 80~dB) speech styles. We evaluated four representative ASR architectures: \textbf{SenseVoice-Small} \cite{an2024funaudiollm}, \textbf{Paraformer} \cite{gao2023funasr}, \textbf{FireRedASR-AED} \cite{xu2025fireredasr}, and \textbf{Whisper} \cite{radford2023robust}. The evaluation encompassed four noise types from the NoiseX-92 database across four SNR levels: $\{5, 0, -3, -5\}$~dB. The experimental results (summarized in Table~\ref{tab:asr_validation}) yield two critical findings. First, across all ASR models and noise conditions, a consistent improvement in recognition accuracy was observed as the Lombard intensity increased from P to L2. Specifically, L2-level speech consistently yielded the lowest word error rates (WER), demonstrating that the acoustic shifts inherent in the Lombard effect are effectively captured by the deep features of modern ASR systems. Second, the performance gap between plain and Lombard speech widens as the SNR decreases. 
This trend confirms that ASR models can provide a discriminative reward signal, particularly in harsh acoustic regimes where intelligibility is compromised. 

Consequently, optimizing for ASR-based rewards provides a practical proxy for inducing the psychoacoustic compensation strategies required for situated conversational agents. Taking into account the trade-off between performance and inference speed, we used SenseVoice-Small to calculate the intelligibility reward during post-training.

\section{Detailed Objective Intelligibility Scores}
This appendix provides the results for the objective intelligibility evaluation visually summarized in Fig.~\ref{fig:object_result} of the main manuscript. Table~\ref{tab:detailed_scores} details the absolute Word Error Rate (WER) and Extended Short-Time Objective Intelligibility (ESTOI) scores for the four evaluated configurations: the vanilla baseline, the RL-only baseline, the SFT-only model, and the proposed NIL-TTS. The evaluation spans six signal-to-noise ratio (SNR) levels, ranging from 6 dB to -9 dB, averaged across four distinct environmental noise types (Babble, Factory, Pink, and White) on both the Seed-TTS \textit{test-zh} and \textit{test-en} datasets. To quantify utterance-level evaluation variance, the table reports the mean absolute scores alongside their corresponding 95\% confidence intervals, computed via paired utterance-level bootstrap resampling.

\begin{DIFnomarkup}
\begin{table}[t]
\centering
\caption{Comparison of basic speech synthesis capabilities under noise-free conditions.}
\label{tab:noise_free_performance}
\small
\renewcommand{\arraystretch}{1.25}

\begin{tabular}{lcccc}
\toprule
\textbf{Model} & \textbf{WER (\%)} $\downarrow$ & \textbf{SIM} $\uparrow$ & \textbf{UTMOS} $\uparrow$ \\
\midrule
\multicolumn{4}{c}{\textbf{Seed-TTS \textit{test-zh}}} \\
\midrule
F5-TTS          & 1.78 & \textbf{0.76} & 2.93  \\
NIL-TTS         & \textbf{1.71} & 0.75 & \textbf{3.15}  \\

\midrule
\multicolumn{4}{c}{\textbf{Seed-TTS \textit{test-en}}} \\
\midrule
F5-TTS          & 1.62 & \textbf{0.68} & 3.68  \\
NIL-TTS         & \textbf{1.50} & 0.67 & \textbf{3.85}  \\

\bottomrule
\end{tabular}
\end{table}
\end{DIFnomarkup}

\section{Basic Speech Synthesis Capabilities Under Noise-Free Conditions}
FiLM operates as a linear modulation mechanism that adjusts the synthesis trajectory via vector scaling and shifting. In our framework, disabling the FiLM branch causes the model to revert to the same inference process as the baseline. We further investigated whether the model’s fundamental speech synthesis performance was compromised by the introduction of Lombard-style guided training.

To evaluate this, we disabled the FiLM input to synthesize standard plain speech. We conducted evaluations on the Seed-TTS \textit{test-zh} set using the following metrics: Word Error Rate (WER, via Paraformer-zh \cite{gao2023funasr}), Speaker Similarity (SIM, based on WavLM-large \cite{chen2022wavlm}), and the objective MOS metric UTMOS \cite{saeki2022utmos}. During inference, we utilized the following hyperparameters: $NFE=32$, $CFG=2.0$, and a sway sampling coefficient of $-1.0$.

As summarized in Table~\ref{tab:noise_free_performance}, the proposed NIL-TTS achieves a lower WER and a higher UTMOS than the F5-TTS baseline, suggesting that the Lombard-style training may even enhance the model's overall intelligibility and naturalness. However, the speaker similarity (SIM) exhibits a marginal decrease from 0.76 to 0.75. We hypothesize that this drop stems from the imperfect decoupling of speaker identity and speaking style within the SSL-based feature representation models such as WavLM \cite{ashihara2024self}. Specifically, as NIL-TTS produces more articulated speech, these stylistic enhancements toward clarity may be perceived by the SIM metric as a deviation from the original speaker's characteristic nuances. Consequently, the metric may mistakenly attribute the shift in speaking style to a reduction in speaker identity.
}


\bibliographystyle{IEEEtran}
\bibliography{refs}

\end{document}